\documentclass[12pt,preprint]{aastex}

\slugcomment{Submitted June 23, 2006}

\shorttitle{AEGIS data sets}
\shortauthors{Davis et~al.}
                                                                                
\begin{document}
                                                                                
\title{The All-wavelength Extended Groth Strip International Survey (AEGIS)
Data Sets}

\author{
M.~Davis\altaffilmark{1},
P.~Guhathakurta\altaffilmark{2},
N.~P.~Konidaris\altaffilmark{2},
J.~A.~Newman\altaffilmark{3,4},
M.~L.~N.~Ashby\altaffilmark{5},
A.~D.~Biggs\altaffilmark{6},
P.~Barmby\altaffilmark{5},
K.~Bundy\altaffilmark{7},
S.~C.~Chapman\altaffilmark{7},
A.~L.~Coil\altaffilmark{3,8},
C.~J.~Conselice\altaffilmark{9},
M.~C.~Cooper\altaffilmark{1},
D.~J.~Croton\altaffilmark{1},
P.~R.~M.~Eisenhardt\altaffilmark{10},
R.~S.~Ellis\altaffilmark{7},
S.~M.~Faber\altaffilmark{2},
T.~Fang\altaffilmark{1},
G.~G.~Fazio\altaffilmark{5},
A.~Georgakakis\altaffilmark{11},
B.~F.~Gerke\altaffilmark{12},
W.~M.~Goss\altaffilmark{13},
S.~Gwyn\altaffilmark{14},
J.~Harker\altaffilmark{2},
A.~M.~Hopkins\altaffilmark{15},
J.-S.~Huang\altaffilmark{5},
R.~J.~Ivison\altaffilmark{6},
S.~A.~Kassin\altaffilmark{2},
E.~N.~Kirby\altaffilmark{2},
A.~M.~Koekemoer\altaffilmark{16},
D.~C.~Koo\altaffilmark{2},
E.~S.~Laird\altaffilmark{2},
E.~Le~Floc'h\altaffilmark{8},
L.~Lin\altaffilmark{2,17},
J.~M.~Lotz\altaffilmark{18,19},
P.~J.~Marshall\altaffilmark{20},
D.~C.~Martin\altaffilmark{21},
A.~J.~Metevier\altaffilmark{2},
L.~A.~Moustakas\altaffilmark{10},
K.~Nandra\altaffilmark{11},
K.~G.~Noeske\altaffilmark{2},
C.~Papovich\altaffilmark{22,8},
A.~C.~Phillips\altaffilmark{2},
R.~M.~Rich\altaffilmark{23},
G.~H.~Rieke\altaffilmark{8},
D.~Rigopoulou\altaffilmark{24},
S.~Salim\altaffilmark{23},
D.~Schiminovich\altaffilmark{25},
L.~Simard\altaffilmark{26},
I.~Smail\altaffilmark{27},
T.~A.~Small\altaffilmark{21},
B.~J.~Weiner\altaffilmark{28},
C.~N.~A.~Willmer\altaffilmark{8},
S.~P.~Willner\altaffilmark{5},
G.~Wilson\altaffilmark{29},
E.~L.~Wright\altaffilmark{23},
and
R.~Yan\altaffilmark{1}.
}

\altaffiltext{1}{Dept.\ of Astron., Univ.\ of California Berkeley, Campbell
  Hall, Berkeley, CA 94720.}
\altaffiltext{2}{UCO/Lick Obs., Univ.\ of California Santa Cruz, 1156 High
  St., Santa Cruz, CA 95064.}
\altaffiltext{3}{Hubble Fellow.}
\altaffiltext{4}{Inst.\ for Nuclear \& Particle Astrophys., Lawrence
  Berkeley Natl.\ Lab., Berkeley, CA 94720.}
\altaffiltext{5}{Harvard-Smithsonian Ctr.\ for Astrophys., 60 Garden St.,
  Cambridge, MA 02138.}
\altaffiltext{6}{UK Astron.\ Technology Ctr., Royal Obs., Blackford Hill,
  Edinburgh EH9~3HJ, United Kingdom.}
\altaffiltext{7}{Dept.\ of Astron., California Inst.\ of Technology, 1201E
  California Blvd., Pasadena, CA 91125.}
\altaffiltext{8}{Steward Obs., Univ.\ of Arizona, 933N Cherry Ave., Tucson,
  AZ 85721.}
\altaffiltext{9}{School of Physics \& Astron., Univ.\ of Nottingham, Univ.\
  Pk., Nottingham NG9~2RD, United Kingdom.}
\altaffiltext{10}{Jet Propulsion Lab., California Inst.\ of Technology,
  4800 Oak Grove Dr., Pasadena, CA 91109.}
\altaffiltext{11}{Imperial College London, Prince Consort Rd., London
  SW7~2BZ, United Kingdom.}
\altaffiltext{12}{Dept.\ of Physics., Univ.\ of California Berkeley,
  Berkeley, CA 94720.}
\altaffiltext{13}{Natl.\ Radio Astron.\ Obs., P.O.~Box\,0, 1003 Lopezville
  Rd., Socorro, NM 87801.}
\altaffiltext{14}{Dept.\ of Physics and Astron., Univ.\ of Victoria,
  Victoria, BC V8W~3P6, Canada.}
\altaffiltext{15}{School of Physics, Univ.\ of Sydney, NSW 2006, Australia.}
\altaffiltext{16}{Space Telescope Science Inst., 3700 San Martin Dr.,
  Baltimore, MD 21218.}
\altaffiltext{17}{Dept.\ of Physics, National Taiwan Univ., Roosevelt Rd.,
  Taipei 106, Taiwan.}
\altaffiltext{18}{Leo Goldberg Fellow.}
\altaffiltext{19}{Natl.\ Optical Astron.\ Obs., 950N Cherry Ave., Tucson, AZ
  85719.}
\altaffiltext{20}{Kavli Inst.\ for Particle Astrophys.\ \& Cosmology, MS29,
  2757 Sand Hill Rd., Menlo Pk., CA 94025.}
\altaffiltext{21}{Space Astrophys., MC~405-47, California Inst.\ of
  Technology, Pasadena, CA 91125.}
\altaffiltext{22}{Spitzer Fellow.}
\altaffiltext{23}{Dept.\ of Physics \& Astron., Univ.\ of California Los
  Angeles, Knudsen Hall, Los Angeles, CA 90095.}
\altaffiltext{24}{Dept.\ of Astrophys., Oxford Univ., Keble Rd., Oxford,
  OX1~3RH, United Kingdom.}
\altaffiltext{25}{Dept.\ of Astron., Columbia Univ.\, 550W 120 St., New York,
  NY 10027.}
\altaffiltext{26}{Assoc.\ of Canadian Univ.\ for Research in Astron.,
  Herzberg Inst.\ of Astrophys., National Research Council, 5071W Saanich
  Rd., Victoria, BC V9E~2E7, Canada.}
\altaffiltext{27}{Inst.\ for Computational Cosmology, Durham Univ., South
  Rd., Durham DH1~3LE, United Kingdom.}
\altaffiltext{28}{Dept.\ of Astron., Univ.\ of Maryland, College Pk., MD
  20742.}
\altaffiltext{29}{Spitzer Science Ctr., California Inst.\ of Technology,
  1200E California Blvd., Pasadena, CA 91125.}

\setcounter{footnote}{29}

\begin{abstract}
In this the first of a series of {\it Letters\/}, we present a description of
the panchromatic data sets that have been acquired in the Extended Groth
Strip region of the sky.  Our survey, the All-wavelength Extended Groth Strip
International Survey (AEGIS), is intended to study the physical properties
and evolutionary processes of galaxies at $z${$\sim$}1.  It includes the
following deep, wide-field imaging data sets:
{\it Chandra\/}/ACIS\footnote{NASA's {\it Chandra\/} X-Ray Observatory was
  launched in July 1999.  The {\it Chandra\/} Data Archive (CDA) is part of
  the {\it Chandra\/} X-Ray Center (CXC) which is operated for NASA by the
  Smithsonian Astrophysical Observatory.} X-ray (0.5--10\,keV),
GALEX\footnote{GALEX (Galaxy Evolution Explorer) is a NASA Small Explorer,
  launched in April 2003.  We gratefully acknowledge NASA's support for
  construction, operation, and science analysis of the GALEX mission,
  developed in cooperation with the Centre National d'Etudes Spatiales of
  France and the Korean Ministry of Science and Technology.} ultraviolet
  (1200--2500\,\AA),
CFHT/MegaCam Legacy Survey\footnote{Based on observations obtained with
  MegaPrime/MegaCam, a joint project of CFHT and CEA/DAPNIA, at the
  Canada-France-Hawaii Telescope (CFHT) which is operated by the National
  Research Council (NRC) of Canada, the Institut National des Science de
  l'Univers of the Centre National de la Recherche Scientifique (CNRS) of
  France, and the University of Hawaii.  This work is based in part on data
  products produced at TERAPIX and the Canadian Astronomy Data Centre as part
  of the CFHT Legacy Survey, a collaborative project of NRC and CNRS.}
  optical (3600--9000\,\AA),
CFHT/CFH12K optical (4500--9000\,\AA),
{\it Hubble Space Telescope\/}/ACS\footnote{Based on GO-10134 program
  observations with the NASA/ESA {\it Hubble Space Telescope\/}, obtained at
  the Space Telescope Science Institute, which is operated by the Association
  of Universities for Research in Astronomy, Inc., under NASA contract
  NAS\,5-26555.} optical (4400--8500\,\AA),
Palomar/WIRC\footnote{Based on observations obtained at the Hale Telescope,
  Palomar Observatory, as part of a collaborative agreement between the
  California Institute of Technology, its divisions Caltech Optical
  Observatories and the Jet Propulsion Laboratory (operated for NASA), and
  Cornell University.} near-infrared (1.2--2.2\,$\mu$m),
{\it Spitzer\/}/IRAC\footnote{This work is based in part on observations made
  with the {\it Spitzer Space Telescope\/}, which is operated by the Jet
  Propulsion Laboratory, California Institute of Technology under a contract
  with NASA.  Support for this work was provided by NASA through contract
  numbers 1256790, 960785, and 1255094 issued by JPL/Caltech.} mid-infrared
  (3.6--8.0\,$\mu$m),
{\it Spitzer\/}/MIPS far-infrared (24--70\,$\mu$m), and
VLA\footnote{The Very Large Array of the National Radio Astronomy Observatory
  is a facility of the National Science Foundation operated under cooperative
  agreement by Associated Universities, Inc.} radio continuum (6--20\,cm).
In addition, this region of the sky has been targeted for extensive
spectroscopy using the DEIMOS spectrograph on the Keck~II 10~m
telescope\footnote{Data presented herein were obtained at the W.\ M.\ Keck
  Observatory, which is operated as a scientific partnership among the
  California Institute of Technology, the University of California, and NASA.
  The Observatory was made possible by the generous financial support of the
  W.\ M.\ Keck Foundation.}.
Our survey is compared to other large multiwavelength surveys in terms of
depth and sky coverage.
\end{abstract}


\section{Introduction}\label{sec:intro}

The All-Wavelength Extended Groth Strip International Survey (AEGIS) is a
collaborative effort to obtain both deep imaging covering all major wavebands
from X-ray to radio and optical spectroscopy over a large area of sky
(0.5--1\,deg$^2$) with the aim of studying the panchromatic properties of
galaxies over the last half of the Hubble time.  The region studied, the 
Extended Groth Strip (EGS: $\alpha$=14$^{\rm h}$17$^{\rm m}$,
$\delta$=+52$^\circ$30$'$) is an extension of and owes its name to a {\it
Hubble Space Telescope\/} ({\it HST\/}) survey consisting of 28 Wide-Field
Planetary Camera\,2 (WFPC2) pointings carried out in 1994 by the WFPC team
\citep*{rho00}.  This field benefits from low extinction, low Galactic and
zodiacal infrared emission, and good schedulability by space-based
observatories, and has therefore attracted a wide range of deep observations
at essentially every accessible wavelength over this comparatively wide
field.

Amongst deep multiwavelength fields, the EGS field provides a unique 
combination of area and depth at almost every waveband observable.  It is two
(for {\it HST\/}) to four (for {\it Spitzer\/} and {\it Chandra\/}) times
larger than the combined GOODS fields \citep{gia04}, yet has a similar range
of wavelength coverage, making it ideal for studying rare classes of objects
that may be absent in smaller fields.  The GEMS field \citep{rix04} covers a
similar area to similar depths, but was studied by the COMBO-17 photometric
redshift survey rather than a spectroscopic survey.  Most AEGIS data sets
cover $\sim$0.5--1\,deg$^2$, considerably smaller than the 2\,deg$^2$ COSMOS
field \citep{koe05}.  However, AEGIS observations are deeper at most
wavelengths, benefiting from greater schedulability and lower backgrounds.
Spectroscopy of the COSMOS field is in progress \citep{lil05}, but will not
be completed for 3--5~years.  An additional advantage of AEGIS is that deep
{\it HST\/}/ACS imaging is available in two bands (F606W and F814W), whereas
the COSMOS field has been observed in F814W only, while the F850LP imaging in
GEMS is too shallow to study subcomponent colors for most galaxies.

Even before AEGIS, the EGS region attracted a rich suite of surveys,
including both spectroscopy \citep{lil95,ste03,cri03} and panchromatic
imaging from both the ground and space, running from X-ray
\citep{miy04,nan05} to ultraviolet (UV) and optical
\citep{bec99,bru99,sar06}, near infrared (IR) \citep{car03,hop00}, mid-IR
\citep{flo99}, submillimeter \citep{cop05}, and radio \citep{fom91}.  AEGIS
has carried this work even further; for example, the first generation of the
DEEP galaxy redshift survey, DEEP1, obtained 620~galaxy redshifts in the
WFPC2 Groth Strip region, now publicly available \citep{sim02,vog05,wei05}.
In comparison, the successor DEEP2 Galaxy Redshift Survey has obtained
9501~redshifts in the EGS so far, with thousands more planned.

Data from DEEP2 are a linchpin for almost all AEGiS studies, providing 
redshifts; internal kinematics for dynamical masses; line strengths for star
formation rates, AGN identification, and gas-phase metallicities; stellar
population ages and metallicites; etc..  The precision and relatively dense
sampling of DEEP2 redshifts allow for accurate measurement of the local
environment of objects in EGS, which is a major factor driving galaxy
evolution.  Other surveys at similar redshifts provide weaker environmental
measures due to larger redshift errors, lower sampling rates, and/or smaller
areas \citep{coo05}.  Furthermore, we can remove cosmic variance fluctuations
from observed AEGIS abundances by comparing redshift distributions to the
other three, widely-separated DEEP2 fields.

Ten instrument teams and a number of theorists are now collaborating on
AEGIS---nearly 100~scientists in half a dozen countries.  The first fruits of
this collaboration are presented in this issue of {\it Astrophysical Journal
Letters\/}.  These papers make use of the power of the combined AEGIS dataset
in a variety of ways.  Six {\it Letters\/} investigate the nature of rare
objects found in the AEGIS field, illustrating the benefits of covering an
area wide enough to find them in \citep{ger06,hua06,kir06,mou06,sym06}.
These studies take advantage of the full multiwavelength coverage from AEGIS,
which provides each object's spectral energy distribution (SED) in detail; we
also explore the range of SEDs exhibited by a wider set of galaxies in
\citet{kon06}.

Five more {\it Letters\/} investigate the drivers and evolution of star 
formation in galaxies using the wide array of indicators available from this
multifaceted dataset, based on UV, IR, and radio continuum as well as optical
emission lines \citep{ivi06,lin06,noe06a,noe06b,wei06}.  Three {\it
Letters\/} focus on the optical properties of Active Galactic Nuclei at
$z${$\sim$}1.4 identified using both deep X-ray and IR data, and explore
their relationship to their large-scale structure environment
\citep{geo06,nan06,pie06}.  Two {\it Letters\/} test for evolution in the
relationships between mass measures to $z${$\sim$}1: between stellar mass and
gas kinematics within individual galaxies \citep{kas06}, and between X-ray
gas emission and galaxy kinematics in groups of galaxies \citep{fan06}.  The
remaining two {\it Letters\/} use DEEP2 spectroscopy to investigate the
nature of objects which are extremely red in optical-IR color, finding that a
substantial fraction of the population lies at $z${$<$}1.4
\citep{con06,wils06}.

These papers present only the first results from the AEGIS survey; we are
just beginning to reach the potential of this manyfaceted dataset.  This {\it
Letter\/} gives details on the AEGIS data which have been obtained so far; we
only describe those survey data sets in the EGS field which are used in this
special issue here.  Large portions of this dataset, including the DEEP2
spectra and {\it HST\/}/ACS imaging, will be publicly released in 2007,
making it a legacy for the entire community.

\section{The Panchromatic Data Sets}\label{sec:data}

\begin{figure} 
\plottwo{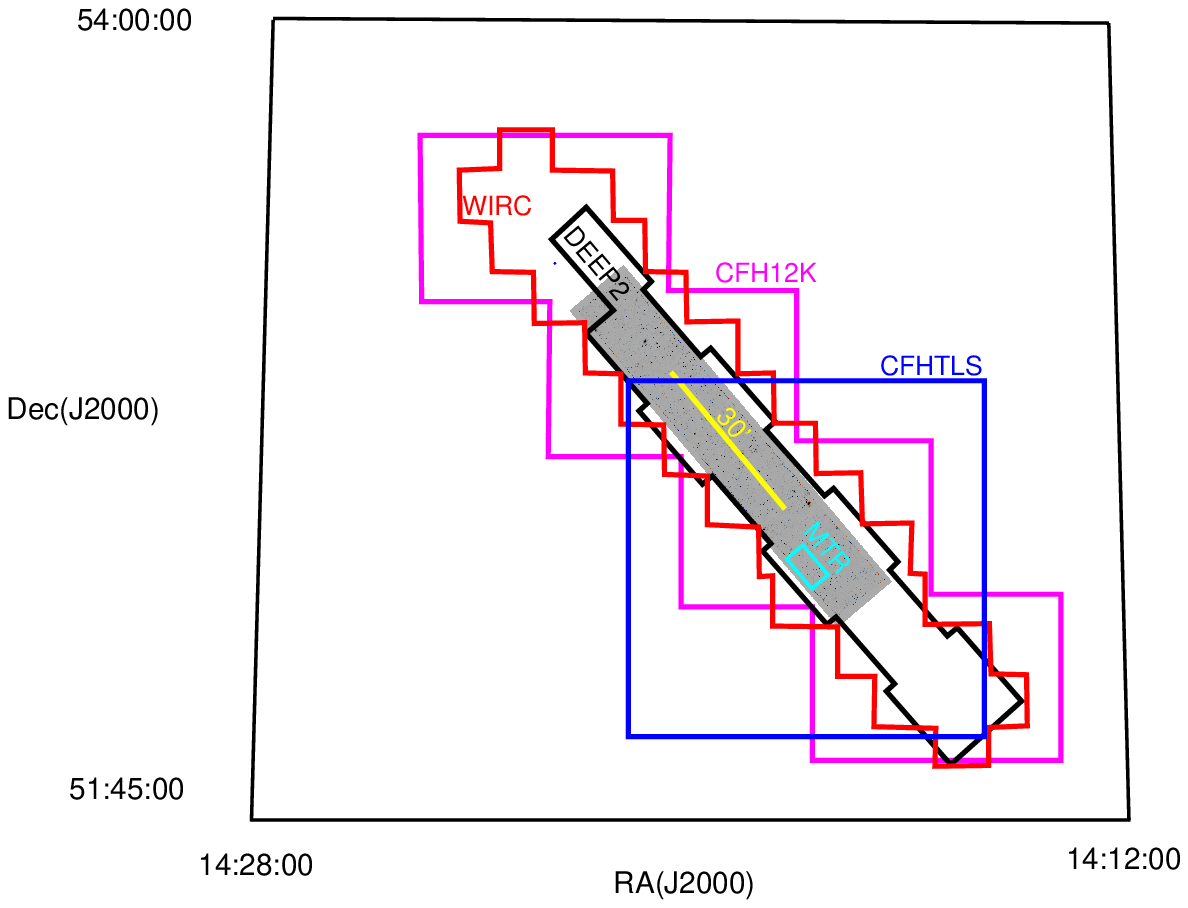}{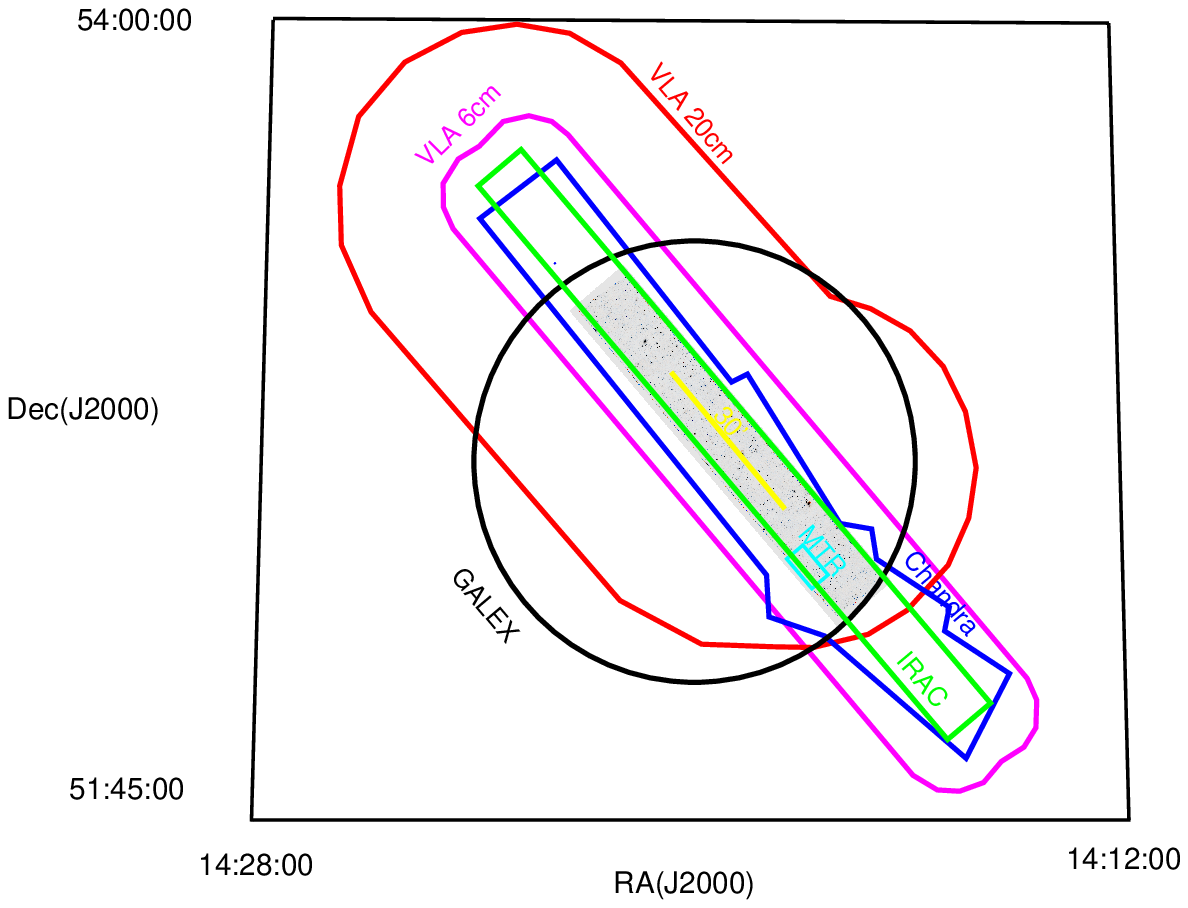}
\caption{Sky coverage maps showing the AEGIS optical and near-IR data sets.
The negative grayscale image shows the {\it HST\/}/ACS mosaic image.  The
outlines represent: CFHT Legacy Survey MegaCam deep optical (dark blue), CFHT
CFH12K optical (pink), Keck DEIMOS spectroscopy (black), and Palomar/WIRC
near-IR (red) on the left, and {\it Chandra\/}/ACIS X-ray (dark blue), GALEX
far- and near-UV (black), {\it Spitzer\/}/IRAC mid-IR and MIPS far-IR
(green), and VLA radio continuum 6\,cm (pink) and 20\,cm (red) on the right.
The length of the {\it HST\/}/ACS mosaic is about 1$^\circ$.  The mini test
region (MTR), shown as a small light blue rectangle superposed on the {\it
HST\/}/ACS image, is used to showcase the panchromatic SEDs of galaxies
\citep{kon06}.  The length of the yellow line segment superposed on the ACS
image running along its length is 30$'$.
}
\label{fig:skymap}
\end{figure}

The basic parameters of the AEGIS multiwavelength data sets are listed in
Table\,\ref{tab:data_tbl} and their sky coverage is shown in
Figure\,\ref{fig:skymap}.  The
acquisition and reduction of the AEGIS data sets and the derivation of source
photometry catalogs [from which panchromatic spectral energy distributions
(SEDs) are measured] are described in detail below.  {\it Spitzer\/} IRS
mid-IR spectra exist for a relatively small number of galaxies; the details
of these observations are in \citet{lef06} and \cite{hua06}.

\begin{deluxetable}{cccccccc}
\rotate
\tablehead{\colhead{Telescope/Instr.} & \colhead{Band} & \colhead{PSF}
& \colhead{$\lambda_{\rm eff}$} & \colhead{Lim.\ Mag.} & \colhead{Surf.\ Dens.}
& \colhead{Area} & \colhead{Exp.\ Time} \\
\colhead{(Mode)} & \colhead{} & \colhead{({\sc fwhm})} & \colhead{} &
\colhead{(5$\sigma$ in most cases)} & \colhead{(deg$^{-2}$)} &
\colhead{(deg$^2$)} & \colhead{(ks)}}
\startdata
{\it Chandra\/}/ACIS  & HB      & 0.5$''$--6.0$''$ & 3.1\,\AA\ (4\,keV) &
     8.2$\times$10$^{-16}\,$erg\,s$^{-1}$\,cm$^{-2}$
& 3200 & 0.67 & 200 \\
     & SB & 0.5$''$--4.0$''$ & 12.4\,\AA\ (1\,keV)&
     1.1$\times$10$^{-16}\,$erg\,s$^{-1}$\,cm$^{-2}$
& 2500 & 0.67 & 200 \\
GALEX    & FUV     & 5.5$''$ & 1539\,\AA\ & 25 (AB) [$3\sigma$] &
8720 & 1.13 & 58 \\
         & NUV     & 5.5$''$ & 2316\,\AA\ & 25 (AB) [$3\sigma$] &
   2.35$\times$10$^4$ & 1.13 & 120 \\
CFHT/MegaCam & $u^*$  & $<$1.1$''$ & 3700\,\AA\ & $\sim$27 (AB) &
   $\approx$10$^5$ & 1 & 6.1 \\
(CFHLS/{\sl current})  & $g'$ & $<$1.0$''$ & 4850\,\AA\ & 28.3 (AB) &
   $\approx$10$^5$ & 1 & 6.5 \\
         & $r'$ & $<$0.9$''$ & 6250\,\AA\   & $\sim$27.5
   (AB) & $\approx$10$^5$ & 1 & 15 \\
	 & $i'$  & $<$0.9$''$ & 7700\,\AA\   & $\sim$27 (AB) &
   $\approx$10$^5$ & 1 & 47 \\
	 & $z'$  & $<$0.9$''$ & 8850\,\AA\   & 26.4 (AB) &
   $\approx$10$^5$ & 1 & 3.6 \\	 
CFHT/CFH12K & $B$   & 1$''$   & 4389\,\AA\   & 24.5 (AB) [8$\sigma$] &
   1.45$\times$10$^5$  & 1.31 & 6.5 \\
         & $R$   & 1$''$   & 6601\,\AA\   & 24.2 (AB) [8$\sigma$] &
   1.45$\times$10$^5$ & 1.31 & 3.6 \\
         & $I$   & 1$''$   & 8133\,\AA\   & 23.5 (AB) [8$\sigma$] &
   1.45$\times$10$^5$ & 1.31 & 4.7 \\
{\it HST\/}/ACS & F606W ($V$)   & 0.1$''$ & 5913\,\AA\   & 28.75
   (AB) [5$\sigma$] & 4.0$\times$10$^5$ & 0.197 & 2.3 \\
(WFC)    & F814W ($I$)   & 0.1$''$ & 8330\,\AA\   & 28.10 (AB) &
3.9$\times$10$^5$ & 0.197 & 2.1 \\
{\it HST\/}/NICMOS & F110W ($J$)   & 0.35$''$& 1.10\,$\mu$m & 25.7 (AB)
   [10$\sigma$] & 3.3$\times$10$^5$ & 0.0128 & 2.6 \\
(NIC3)  & F160W ($H$) & 0.35$''$& 1.59\,$\mu$m & 25.5 (AB)
   [10$\sigma$] & 3.3$\times$10$^5$ & 0.0128 & 2.6 \\
Palomar/WIRC  & $J$ & 1$''$   & 1.25\,$\mu$m & 23 (Vega) &
   7.64$\times$10$^4$ & 0.2 & 18 \\
	 & $K_s$  & 1$''$   & 2.14\,$\mu$m  & 20.6 (Vega) &
   5.37$\times$10$^4$ & 0.7 & 11 \\
{\it Spitzer\/}/IRAC & Band\,1  & 1.8$''$ & 3.6\,$\mu$m & 0.9\,$\mu$Jy &
   1.66$\times$10$^5$ & 0.33 & 10.1 \\
      	 & Band\,2  & 2.0$''$ & 4.5\,$\mu$m & 0.9\,$\mu$Jy &
   1.68$\times$10$^5$ & 0.33 & 10.1 \\
      	 & Band\,3  & 2.2$''$ & 5.8\,$\mu$m & 6.3\,$\mu$Jy & 
   4.90$\times$10$^4$ & 0.33 & 10.1 \\
      	 & Band\,4  & 2.2$''$ & 8.0\,$\mu$m & 5.8\,$\mu$Jy &
   4.86$\times$10$^4$ & 0.33 & 10.1 \\
{\it Spitzer\/}/MIPS & 24\,$\mu$m & 5.9$''$  & 23.7\,$\mu$m & 77\,$\mu$Jy &
   1.76$\times$10$^4$ & 0.534 & 1.68 \\
	 & 70\,$\mu$m & 19$''$   & 71.4\,$\mu$m & 10.3\,mJy & 
   1275 & 0.498 & 0.84 \\
VLA	 & 6\,cm     & 1.2$''$ & 6\,cm       & 0.55\,mJy\,beam$^{-1}$
   [10$\sigma$] & 88.9 & 5.2 & 0.5 \\
   	 & 20\,cm    & 4.2$''$ & 20\,cm  & 100\,$\mu$Jy\,beam$^{-1}$ &
   1075 & 0.64 & 400 \\
\hline
\hline
Telescope/Instr. & Wavelen.\ Range & Spec.\ Res. & Spatial PSF &
Lim.\ Mag. & Area & Num.\ Targ. & Exp.\ Time \\
(Mode) & (\AA) & (\AA) & ({\sc fwhm}) &  & (deg$^2$) &  & (ks) \\
\hline
Keck/DEIMOS & 6400--9100 & 1.4 & 0.6$''$--1$''$ & $R_{\rm AB}$=24.1
& 0.5 & 17,600 & 3.6 \\
(DEEP2/Spectra) &  &  &  &  & ({\sl final}) & ({\sl final}) &
\enddata
\label{tab:data_tbl}
\end{deluxetable}

\subsection{Chandra ACIS X-ray Images}\label{sec:chandra}

The EGS region has been surveyed at X-ray wavelengths by {\it Chandra\/}
using the Advanced CCD Imaging Spectrometer (ACIS).  The observations consist
of eight~individual pointings obtained with the ACIS-I instrument with a
pixel scale of $\approx$0.49$''$ and a field of view of
$\approx$17$'${$\times$}17$'$.  The total exposure time per pointing is
$\sim$200\,ks split into at least four separate integrations obtained at
different epochs.

The data reduction is performed using the {\sc ciao} data analysis software
version\,3.2.  After identifying hot pixels and cosmic-ray afterglows, the
raw data are processed through the {\sc ciao} {\sc acis\_process\_events}
task, which applies the latest gain map and charge transfer inefficiency
corrections.  The observations are then screened using the standard ASCA
grade set (event grades 0, 2, 3, 4, and 6) and inspected for flaring spikes,
when the background deviates by more than 3$\sigma$ from the mean quiescent
value.  Bad time intervals amount to $<5$\% of the total exposure time per
pointing.  Individual observations of a given {\it Chandra\/} pointing are
then merged into a single event file. 

The final mosaic images are used to detect sources in a number of energy
bands, including 0.5--2.0\,keV, 2.0--7.0\,keV, and 0.5--7.0\,keV.  The source
detection is based on pre-selection of candidate sources using the {\sc ciao}
{\sc wavdetect} task with a low likelihood probability threshold ($10^{-4}$),
followed by aperture count extraction using the 70\% PSF radius and a local
background determination to estimate the source significance \citep[for
details see][]{nan05}.  The final catalog comprises sources with Poisson
false probability $<$4$\times$10$^{-6}$.  Point-source fluxes are estimated
by integrating the net counts within the 90\% encircled energy radius at the
position of the source.  We adopt a power-law SED with $\Gamma$=1.4 and
Galactic neutral hydrogen column density $N_{\rm
H}$=1.3$\times$10$^{20}$\,cm$^{-2}$, appropriate for the EGS field.  The
limiting fluxes in standard bands are listed in Table\,\ref{tab:data_tbl}.
Only a subset of the AEGIS {\it Chandra\/} observations has been analysed at
this point and therefore used for followup analysis \citep[see][]{geo06}.

At the time of writing the AEGIS {\it Chandra\/} observations represent the
third deepest X-ray survey in the sky.  The Hubble Deep Field North survey is
$\sim$5$\times$ deeper over a $\sim$5$\times$ smaller area \citep{ale03}.
The {\it Chandra\/} Deep Field South \citep{gia02,leh05} covers about half
the AEGIS survey area mostly to a similar depth, with a smaller central
region that is twice as deep.  The vast majority of X-ray sources detected in
the EGS field are active galactic nuclei (AGN).  At the target redshift of
the DEEP2 survey, $z$=1, the AEGIS limiting luminosity of
7$\times$10$^{41}$\,erg\,s$^{-1}$ corresponds to $\sim$0.005$\times${$L_*$},
where $L_*$ is the characteristic AGN luminosity at this redshift
\citep{bar05}. 

\subsection{GALEX Ultraviolet Images}\label{sec:galex}

The Galaxy Evolution Explorer (GALEX) images a 1.25$^\circ$ diameter field.
GALEX images of the EGS are built from stacks of 42 far-ultraviolet (FUV) and
87 near-ultraviolet (NUV) separate one-orbit images obtained in 2003, 2004,
and 2005 and processed using version~4.1 of the GALEX pipeline.  The combined
exposure times are 5.8$\times$10$^4\,$s for FUV and 1.2$\times$10$^5\,$s for
NUV.  The raw photon count images are flat-fielded and calibrated using
relative response maps.  The resulting calibrated intensity images are in
units of photon\,s$^{-1}$, where 1\,photon\,s$^{-1}$ corresponds to 18.82 and
20.08 AB mag for the FUV and NUV bands, respectively.  The background in the
images is estimated using PoissonBG, a program written for GALEX data that
uses Poisson rather than Gaussian statistics to clip suspected sources from
the background map.  Finally, the source catalogs are derived from the
background-subtracted images with SExtractor \citep{ber96}.
More details about the GALEX pipeline
can be found at {\tt http://www.galex.caltech.edu}.

\subsection{Canada-France-Hawaii Telescope Legacy Survey MegaCam Optical
Images}\label{sec:cfhls}

The EGS is one of four 1\,deg$^2$ fields covered by the ongoing
Canada-France-Hawaii Telescope Legacy Survey (CFHTLS) Deep
Survey\footnote{\tt
http://www.cfht.hawaii.edu/Science/CFHLS/cfhtlsdeepwidefields.html}.  This
field (labelled ``D3'' by CFHTLS) has been observed for a total of 114\,hr
using the MegaCam imager on the 4\,m Canada-France-Hawaii Telescope
\citep[CFHT;][]{bou03} from April 2003 to the present; a total integration
time of 330\,hr is planned by the end of the five-year survey.  This time is
divided amongst five broad-band filters: $u^*$, $g'$, $r'$, $i'$, and $z'$.
Only images with seeing {\sc fwhm} smaller than 0.9$''$--1.1$''$ (depending
on the band) are included in the survey.  Five-sigma point source detection
limiting AB magnitudes in the current data set range from 26.4 (in $z'$) to
28.3 (in $g'$).

The catalogs used in AEGIS papers are based on a subset of the CFHTLS data
set with total exposure times of 1.7/1.8/4.1/13.0/1.0\,hr in
$u^*$/$g'$/$r'$/$i'$/$z'$.  After visual rejection of defective exposures,
Elixir-processed frames were run through the {\sc AstroGwyn} and {\sc
PhotGwyn} software packages, improving both astrometric and photometric
calibrations significantly \citep{mag04,gwy06}; they were then corrected for
distortions and coadded using {\sc SWarp}\footnote{\tt
http://terapix.iap.fr/rubrique.php?id\_rubrique=49}.  Photometry was then
obtained using the double image mode of SExtractor \citep{ber96}, with $i'$
used as reference image.  The principal CFHTLS measurements used here are
Kron aperture (SExtractor {\tt MAG\_AUTO}) AB magnitudes.  In addition,
photometric redshifts for CFHTLS sources were determined by applying the {\sc
gwynz} code to photometry within matched, 1$''$ radius apertures
\citep{gwy06}.  Additional details of the procedures used are available at:
{\tt http://www.astro.uvic.ca/grads/gwyn/cfhtls/D3.html}.

\subsection{Canada-France-Hawaii Telescope CFH12K Optical
Images}\label{sec:cfh12k}

The EGS was imaged in $B$, $R$, and $I$ bands using the CFH12K mosaic camera
\citep{cui01} on CFHT.  This 12K$\times$8K mosaic camera has a scale of
0.21$''$\,pixel$^{-1}$ and a field of view of
0.70$^\circ${$\times$}0.47$^\circ$ (with the longer axis oriented East-West).
The $R$-band seeing ranged from 0.75$''$--1$''$ {\sc fwhm} in the four
separate CFHT12K pointings covering the EGS; integration times were
$\sim$1\,hr in $B$ and $R$ and $\sim$2\,hr in $I$.  The $R$-band magnitudes
were measured within circular apertures of radius 3$r_g$, where $r_g$ is the
$\sigma$ of a Gaussian fit to the image profile; for objects where
3$r_g${$<$}1, a 1$''$ radius aperture was used instead.  The $B${$-$}$R$ and
$R${$-$}$I$ colors of each object were measured using a 1$'"$ radius
aperture.  The resulting $BRI$ photometry was calibrated to the AB system
within the native CFHT12K passbands (which differ significantly from the
Kron-Cousins system, particularly in $I$) using stars observed by the Sloan
Digital Sky Survey \citep[SDSS;][]{yor00}; the $BRI$ stellar locus is used to
ensure consistency of the color system between CFH12K pointings.  Details of
the data reduction, astrometry, star-galaxy separation, and catalog
construction can be found in \citet{coi04}; the resulting catalogs are
available at {\tt http://deep.berkeley.edu/DR1}. 

\subsection{Hubble Space Telescope ACS Optical Images}\label{sec:acs}

Deep {\it HST\/} images of the EGS were obtained with ACS as part of GO
Program~10134 (PI: M.~Davis).  The EGS was imaged in the $V$ (F606W) and $I$
(F814W) bands during the period 2004 June to 2005 March.  A mosaic pattern
consisting of 21$\times$3\,=\,63 contiguous ``tiles'' was used to cover an
effective area of $\sim$10.1$'${$\times$}70.5$'$\,=\,710.9\,arcmin$^2$
following the IRAC imaging strip (\S\,\ref{sec:irac}).  The exposure times
per ``tile'' were 2260 and 2100\,s in the $V$ and $I$ bands, respectively.
Tiles were observed at a position angle of 130$^\circ$, or rotated by
multiples of 90$^\circ$ relative to this value to meet guide star
constraints.  Each tile was observed in a 4-pointing dither pattern in each
filter, in order to achieve half-pixel dithering at the center of ACS WFC,
bridge the detector gap, and improve tile overlap.  The final mosaic is
gap-free and each pixel is observed at least three times.  Dithered pointings
were combined with the {\sc stsdas} multidrizzle package using a square
kernel.  The final images have a pixel scale of 0.03$''$\,pixel$^{-1}$ with a
PSF of 0.12$''$ {\sc fwhm}.  The 5$\sigma$ limiting magnitudes for a point
source are $V_{\rm F606W}=28.14$ (AB) and $I_{\rm F814W}=27.52$ (AB) within a
circular aperture of radius 0.12$''$ ($\sim$50\,pixel area).  For an extended
object, the 5$\sigma$ limiting magnitudes are $V_{\rm F606W}=26.23$ (AB) and
$I_{\rm F814W}=25.61$ (AB) for a circular aperture of radius $0.3''$
($\sim$314\,pixel area).

We detected objects in summed ACS $V$+$I$ images and constructed initial
galaxy segmentation maps using the SExtractor galaxy photometry software
\citep{ber96} and a detection threshold of 1.5$\sigma$ and 50\,pixels.  These
detection maps and the ACS zeropoints \citep{sir05} were applied to each band
separately to create the ACS photometric catalogs.

\subsection{Keck DEIMOS Optical Spectra}\label{sec:deimos}

The EGS is one of the four fields observed by the DEEP2 collaboration
\citep{dav03}.  Here we briefly describe the DEEP2 data in EGS; for more
details see \citet{dav05} for maskmaking algorithms, Faber et~al.\ (in prep.)
for full survey details, and Cooper et~al.\ (in prep.) for data reduction
pipelines.  Targets were selected for DEEP2 spectroscopy from the CFHT12K
$BRI$ imaging described in \S\,\ref{sec:cfh12k}.  Eligible DEEP2 targets have
18.5$\le${$R$}$\le$24.1, $>$20\% probability of being a galaxy (based on
angular size, $B${$-$}$R$/$R${$-$}$I$ colors, and $R$ mag), and surface
brightness brighter than:
\begin{equation}
\mu_R=R+2.5{\rm log}_{10}(A)\leq26.5,
\end{equation}
where $A$ is the area of the aperture used to measure the CFHT12K $R$
magnitude (\S\,\ref{sec:cfh12k}); all magnitudes are AB.  Each object is
given a weight based on its probability of being a galaxy, its $R$ magnitude,
and whether or not meets the DEEP2 color cut used to eliminate low-$z$
objects in other fields [galaxies with
($B${$-$}$R$)\,$<$\,2.45($R${$-$}$I$)$-$0.5, ($R${$-$}$I$)$>$1.1, or
($B${$-$}$R$)$<$0.5 all pass this cut].  This weight is used when randomly
selecting amongst multiple objects that cannot be observed simultaneously due
to DEIMOS slitmask constraints \citep{dav05}.  Fainter objects (particularly
those with $R${$>$}21.5 and expected $z${$<$}0.75 from the color cut) are
given lower weight in order to sample a range of luminosities and roughly
equal numbers of galaxies below and above $z${$=$}0.75 (Faber et~al.\ in
prep.).  Selection probabilities for each potential target are known to
$<$1\%; the median is $>$70\% for objects with $z${$<$}0.1, falls to 54\% at
$z${$=$}0.5--0.6, and is flat at 73\% for $z${$>$}0.8.

All spectra were taken with the DEIMOS spectrograph \citep{fab03} at the
Keck\,II\ telescope.  Each observation uses a unique aluminum mask milled
with $\sim$150 1$''$ wide and $>$3$''$ long slitlets over a
16$'${$\times$}4$'$ area, which is observed for a minimum of 1\,hr (until a
target signal-to-noise is reached) divided amongst 3 or more sub-exposures.
The 1200\,line\,mm$^{-1}$ grating used yields a dispersion of
$\sim$0.33\,\AA\,pixel$^{-1}$ and a spectral resolution of {\sc
fwhm}=1.4\,\AA.  The typical wavelength range of the spectra is
$\sim$6500--9100\,\AA, varying modestly with slit position.  Slitlets are
tilted up to 30$^\circ$ to follow the photometric major axes of extended
targets.

All DEEP2 spectra were reduced with an IDL pipeline heavily modified from the
{\sc idlspec2d} package designed for SDSS (Burles \& Schlegel in prep.).
Spectra were extracted using both boxcar and optimized Gaussian weighting.
The pipeline determines a set of candidate redshifts for each object by
fitting a linear combination of templates at each possible $z$ and finding
local minima in $\chi^2(z)$; these redshifts are then evaluated and selected
amongst by DEEP2 team members using a graphical interface.  Two significant
features must match the templates for a secure redshift (quality $z_q$=3 or
4); a resolved [O\,{\sc ii}] 3727\,\AA\ doublet is counted as two features.

Based on both repeated observations and tests with multiwavelength
photometry, we estimate that $\lesssim$5\% of $z_q$=3 redshifts (obtained for
11\% of EGS targets) are incorrect, while $\lesssim$0.5\% of
highest-confidence, $z_q$=4 redshifts (60\% of EGS targets) are incorrect.
Lower quality redshifts are considered insecure or ambiguous and not used for
any analyses.  Objects with repeated observations (310 out of the 13570
galaxies observed so far in EGS) have an rms redshift uncertainty of
30\,km\,s$^{-1}$.  Secure redshifts have been obtained for 9501 galaxies in
EGS, with median redshift 0.74.  Objects at $z${$>$}1.42 tend not to have
strong features in the DEEP2 spectral window; such objects appear to comprise
the bulk of DEEP2 redshift failures (C.~Steidel, priv.\ comm.).

\subsection{Hubble Space Telescope NICMOS Near-Infrared
Images}\label{sec:nicmos}

Deep {\it HST\/} Near Infrared Camera and Multi-Object Spectrometer (NICMOS)
NIC3 images of the EGS were obtained as ``parallels'' of the ACS images
(\S\,\ref{sec:acs}) as part of GO Program\,10134 (PI: M.\,Davis).  Each NIC3
field was observed in the $J$ (F110W) and $H$ (F160W) bands for 2560\,s per
band (one {\it HST\/} orbit).  The field of view of a single NIC3 pointing is
51.2$''${$\times$}51.2$''$.  The resulting 63\,NIC3 fields cover a combined
area of $\approx$46\,arcmin$^2$.  Out of the 63, 58 fully overlap with the
ACS imaging mosaic; the remaining five NIC3 pointings coincide with other
AEGIS data sets.  Since the NIC3 PSF is undersampled, we developed a 4-point
dither pattern that simultaneously provides optimal sub-pixel dithering for
ACS WFC and NIC3 to improve the final resolution of the reduced images.

The NICMOS NIC3 images were processed in similar way as the ACS images.
Basic image reductions were performed with the {\sc stsdas.calnica} routine:
flatfielding and corrections for dark current, bias, variable quadrant bias,
amplifier glow, cosmic ray persistence, detector nonlinearity, pixel defects,
bad imaging regions, cosmic rays of unusual size, and the count-rate
dependent nonlinearity.  The four pointings per tile per filter were combined
with the {\sc stsdas.multidrizzle} package using a square kernel.  The final
images have a scale of 0.1$''$\,pixel$^{-1}$ with a PSF of 0.35$''$ {\sc
fwhm}.  Sources were detected using summed NIC3 $J$+$H$ images using
SExtractor \citep{ber96}, using a detection threshold of 1$\sigma$ and
minimum size of 10\,pixels.  Photometry was performed on the individual
images using circular apertures of diameter 0.52$''$ calibrated using the
zero-points in the NICMOS Data Handbook.

\subsection{Palomar WIRC Near-Infrared Images}\label{sec:wirc}

Near-IR observations of the EGS in the $J$ and $K$ bands were obtained using
the Wide-field Infrared Camera (WIRC) on the Palomar 5\,m telescope.  The
observations were carried out between 2003 and 2005.  WIRC has an effective
field of view of 8.1$'${$\times$}8.1$'$, with a scale of
0.25$''$\,pixel$^{-1}$.  The EGS observations consist of 33~WIRC overlapping
pointings in $K$ and 10~pointings in $J$, each with 4$\times$30\,s exposures
dithered over a non-repeating 7$''$ pattern.  Typical total exposure times
per band at any given location within the EGS are 1--2\,hr.  The net seeing
{\sc fwhm} ranges from 0.8$''$ to 1.2$''$.  Photometric calibration was
carried out by referencing standard stars during photometric conditions.  The
final images were made by combining individual mosaics obtained over several
nights.  The images were processed using a double-pass reduction pipeline
developed specifically for WIRC.  The total area covered in the $K$ band is
2400\,arcmin$^2$=0.67\,deg$^2$, with about a third of this area covered in
the $J$ band.

\subsection{Spitzer IRAC Mid-Infrared Images}\label{sec:irac}

The {\it Spitzer} mid-IR observations were carried out as part of Guaranteed
Time Observing (GTO) program number 8, using time contributed by G.~Fazio,
G.~Rieke, and E.~Wright.  The Infrared Array Camera \citep[IRAC;][]{faz04}
observations were performed in two epochs, 2003 December and 2004 June/July.
Each IRAC exposure covered a 5.12$'${$\times$}5.12$'$ field of view with
256$\times$256 pixels and a scale of 1.2$''$\,pixel$^{-1}$.  At each of
52~positions in a 2$^\circ${$\times$}10$'$ map there were dithered 200\,s
exposures at 3.6, 4.5, and 5.8\,$\mu$m, together with 208~dithered 50\,s
exposures taken concurrently at 8.0\,$\mu$m.

Data processing began with the Basic Calibrated Data produced by version\,11
of the Spitzer Science Center IRAC pipeline.  Individual frames were
corrected for the `muxbleed' artifact near bright stars.  Mosaicing was done
using custom IDL scripts: each individual frame was distortion-corrected and
projected onto a reference frame, and the frames were combined by averaging
with 3$\sigma$-clipping.  Rejection of cosmic rays, scattered light, and
other image artifacts was accomplished by the sigma-clipping during mosaicing
and also facilitated by having the observations done at two position angles
differing by $\sim$180$^\circ$.  The scale of the mosaics,
0.6$''$\,pixel$^{-1}$, sub-samples the native IRAC pixel scale by a factor of
two.  The two shorter-wavelength IRAC bands are more sensitive than the
longer-wavelength bands.

To make catalogs, sources in the IRAC mosaics were identified using {\sc
daophot/find}, and photometered in a 3$''$ diameter aperture.  Aperture
corrections to the IRAC calibration photometry aperture of 12.2$''$
(multiplicative factors of 2.07, 2.15, 2.45, and 2.68 in the four~bands) were
applied.  For each object, neighboring objects in a 200-pixel box were
subtracted before photometry.  Both 3.6\,$\mu$m- and 8.0\,$\mu$m-selected
catalogs were generated: photometry for the objects in each was centered on
the position in the mosaic in the selected band.  There are about
73,000~objects in the 3.6\,$\mu$m-selected catalog, many of which are
undetected in the 8.0\,$\mu$m mosaic; the 8.0\,$\mu$m catalog contains only
16,000~objects.

\subsection{Spitzer MIPS Far-Infrared Images}\label{sec:mips}

{\it Spitzer} far-IR observations with the MIPS instrument \citep{rie04} were
also carried out as part of the same GTO program.  Data were obtained in
January and June 2004 using the slow rate MIPS scan mode with legs
2.4$^\circ$ long.  The MIPS 24\,$\mu$m channel has a 5.4$'${$\times$}5.4$'$
field of view (128$\times$128 array of 2.55$''$ pixels).  The 70\,$\mu$m
channel has a 5.2$'${$\times$}5.2$'$ field of view (32$\times$32 array of
9.98$''$ pixels), but only half of the array is functional.  The 160\,$\mu$m
channel has a 5.3$'${$\times$}2.1$'$ field of view (20$\times$3 array of
16$''${$\times$}18$''$ pixels), but one row of the array is not operational.
The final mosaic covers an area $\sim$2.4$^\circ${$\times$}10$'$.  The
effective integration time at 24\,$\mu$m is $\sim$1500\,s for locations near
the long centerline of the strip, decreasing to $\sim$700\,s 5$'$ from the
centerline.  At 70 and 160\,$\mu$m, the average integration times are
$\sim$700 and 100\,s\,pixel$^{-1}$ respectively.  The data were reduced and
mosaiced with the MIPS Data Analysis Tool \citep{gor05}.  Sources were
identified and photometry extracted with PSF fitting using the {\sc daophot}
software \citep{ste87}.

\subsection{VLA Radio Continuum Images}\label{sec:vla}

Radio continuum observations at 6\,cm (4.8\,GHz, C~band) were obtained at the
Very Large Array (VLA) in BnA configuration for a total of 19\,hr during 2003
October 11--13 (the continuum mode maximizes sensitivity but decreases the
effective field of view).  At 4.8\,GHz the VLA antennas have a primary beam
{\sc fhwm} of 9$'$.  The mapping grid contained 74~pointings, spaced 5$'$
apart, providing roughly uniform sensitivity over a 17$'${$\times$}2$^\circ$
strip.  Each pointing center was observed for 15\,min.  Phase stability was
sufficient to yield astrometric accuracy better than 0.1$''$ rms, and flux
density was calibrated relative to 3C\,286.  Images containing bright
($>$10\,mJy\,beam$^{-1}$) point sources were self-calibrated.

The data were reduced using the {\sc aips} software package.  For each of the
74~pointings, we created a 2048$\times$2048\,pixel image with a scale of
0.4$''$\,pixel$^{-1}$.  To avoid clean bias we {\sc clean}-ed the images to a
flux level of 260\,$\mu$Jy\,beam$^{-1}$, corresponding to
$\sim$4$\sigma$~rms, which took $\sim$100--200~iterations.  For each
quarter of the length of the strip, overlapping images were combined into a
mosaic using the {\sc linmos} task in the {\sc miriad} software package.
The rms noise in the mosaiced images is 42\,$\mu$Jy\,beam$^{-1}$.  Source
candidates were extracted using SExtractor \citep{ber96} and photometry was
performed with {\sc jmfit} in {\sc aips} on the individual pointing images,
with corrections for delay beam distortions.

Altogether 51 radio components (some of which may be double radio sources)
were detected at $\ge$10$\sigma$ significance.  Further details are in
Table~1, and the source list and identification with IRAC and DEEP2
counterparts are given by \citet{will06}.

Observations at 20\,cm (1.4\,GHz, L~band) are described by \citet{ivi06}
where the resulting catalog, AEGIS20, containing 1123~discrete radio emitters
is published in electronic form.  Briefly, spectral-line data were obtained
using the VLA in B configuration, with correlator mode~`4', for a total of
110\,hr in 2003 December and 2005 April--June.  Eight overlapping pointings
were observed spanning the length of the EGS, concentrating on the six
furthest from the bright source, 3C\,295.  For each pointing the $\sim$30$'$
{\sc fwhm} primary beam was blanketed with 37~images, each comprising
512$\times$512\,pixels with a scale of 0.8$''$\,pixel$^{-1}$, with
10--20~additional images centered on more distant, bright sources (including
3C\,295).  Central images of the six main pointings were mosaiced, correcting
for the primary beam response and excluding data beyond the half-power point.
The resulting image covers 0.04, 0.36, and 0.64\,deg$^2$ to $5\sigma$ limits
of 50, 75, and 100\,$\mu$Jy\,beam$^{-1}$.  Source detection for AEGIS20
followed that described by \citet{big06}.

\section{Summary}\label{sec:summy}

This {\it Letter\/} has described the multiwavelength dataset in the Extended
Groth Strip assembled by the AEGIS collaboration.  The remaining {\it
Letters\/} in this special issue discuss a wide range of scientific results 
derived from these data.  By combining deep observations at almost every
wavelength available over $\sim 1$ deg$^2$ of sky with the relatively high
resolution spectroscopy and dense sampling of the DEEP2 Galaxy Redshift
Survey, AEGIS is making possible many unique studies of the evolution of 
galaxies over more than half the history of the universe.  The high-quality
internal kinematics and environment measurements in EGS are unmatched amongst
deep multiwavelength fields.   The dataset is continuing to grow and its
potential is only beginning to be tapped; we expect AEGIS to provide a legacy
long into the future.

\acknowledgments

This research has made use of NASA's Astrophysics Data System Bibliographic
Services.  The authors wish to recognize and acknowledge the very significant
cultural role and reverence that the summit of Mauna Kea has always had
within the indigenous Hawaiian community.  We are most fortunate to have the
opportunity to conduct observations from this mountain.  ALC and JAN are
supported by NASA through Hubble fellowship grants HF-01182 and HF-01165
awarded by STScI, which is operated by AURA, Inc., for NASA, under contract
NAS 5-26555.  JML acknowledges support from the NOAO Leo Goldberg Fellowship,
NASA/STScI grants GO-10134 and AR-10675, NASA NAG5-11513 grant to P.~Madau,
and a Calspace grant to D.~C.~Koo.  LAM's work was carried out at
JPL/Caltech, under a contract with NASA.  SAK would like to thank Eddie
Bergeron for assistance with reducing and calibrating the NICMOS data.

\end{document}